\documentclass[12pt]{article}

\usepackage[left=2.8cm,right=2.8cm,top=2.8cm,bottom=2.8cm]{geometry}

\usepackage[colorlinks=true,linkcolor=black,citecolor=black,urlcolor=black,
filecolor=black]{hyperref}

\usepackage{latexsym}
\usepackage{amsmath}
\usepackage{amsfonts}
\usepackage{mathrsfs}
\usepackage{dsfont}
\usepackage{verbatim}
\usepackage[nosort]{cite}

 \csname
@addtoreset\endcsname{equation}{section}

\def\a{\alpha}

\def\d{\delta}

\def\e{\epsilon}

\def\L{\Lambda}

\def\p{\pi}

\def\vf{\varphi}

\def\be{\begin{equation}}
\def\ee{\end{equation}}
\def\bea{\begin{eqnarray}}
\def\eea{\end{eqnarray}}
\def\ba{\begin{array}}
\def\ea{\end{array}}

\def\12{\frac{1}{2}}

\begin{document}
\vspace*{-2cm}
\begin{flushright}
\begin{tabular}{c}
AEI-2013-224
\end{tabular}
\end{flushright}

\vspace*{1pt}

\begin{center}
{\Large\sc On the higher-spin charges of conical defects}

\vspace{15pt}
{\sc A.~Campoleoni$^{1}$ and S.~Fredenhagen$^{2}$}

\vspace{20pt}
{\sl\small

$^{1}$\begin{minipage}[t]{7.5cm}Universit{\'e} Libre de Bruxelles\\
and International Solvay Institutes\\
ULB-Campus Plaine CP231\\
1050 Brussels,\ Belgium\\[5pt]
\texttt{andrea.campoleoni@ulb.ac.be}
\end{minipage}
 $^{2}$\begin{minipage}[t]{7.5cm}Max-Planck-Institut f{\"u}r Gravitationsphysik\\
Albert-Einstein-Institut\\
Am M{\"u}hlenberg 1\\
14476 Golm,\ Germany\\[5pt]
\texttt{stefan.fredenhagen@aei.mpg.de}
\end{minipage}}

\vspace{30pt}{\sc\large Abstract}\end{center}

The conical defect solutions in higher-spin gauge theories on 2+1
dimensional space-times with AdS-asymptotics are conjectured to
correspond to certain primary fields in the dual conformal field
theory on the boundary. In this note we prove that indeed all
higher-spin charges match. 

\vspace{10pt}



\section{Introduction}\label{sec:intro}

When the dimension of space-time is equal to three, one can build a
variety of higher-spin gauge theories. A notable class admits a
Chern-Simons formulation with gauge algebra $sl(N,\mathbb{R}) \oplus
sl(N,\mathbb{R})$ \cite{general3D}, and models the interactions of a set of
tensors of rank $2,3,\ldots,N$. In this context we investigate the
solutions introduced in \cite{conical}, whose metric displays a
conical singularity in a particular gauge. These backgrounds
approach asymptotically $AdS_3$ in the sense described in
\cite{HR,spin3}, and play an important role in the conjectured
holographic duality between $W_N$ minimal models and Vasiliev theories
\cite{GG}.\footnote{See also \cite{dual-review} for a recent review,
although the proposal for the boundary dual of the solutions we
consider was later refined in \cite{conical-CFT}.} The latter are
extensions of the Chern-Simons setup that include matter couplings
\cite{VP}, but the conical defects can be considered as solutions of
the Vasiliev equations in which the scalars vanish. They are
conjectured to correspond to specific primary states in the minimal
model dual \cite{conical,triality,conical-CFT}. 

A strong evidence in support of this identification is the matching of higher-spin charges on both sides of the duality. However, up to
present this check was performed explicitly only for the first few
charges. In the following we fill this gap, and show that all charges match in the semi-classical regime.

Our proof goes as follows: in the higher-spin gauge theory, one can
compute straightforwardly the charges in the so-called $u$-basis. On
the other hand, it is precisely the $u$-basis for which one has a
free-field construction in the boundary conformal field theory by the
(quantum) Miura transformation.  To compare to the higher-spin theory,
we have to map the standard Miura transformation from the plane
to the cylinder by a conformal transformation, because in the
higher-spin theory we consider solutions defined on manifolds
with the same topology as $AdS_3$, whose boundary is a cylinder. The
free-field construction of the $u$-currents on the cylinder then makes
it possible to compute their zero-mode eigenvalues on ground states,
which precisely match the $u$-charges of the conical solutions.

\section{Conical defects and their higher-spin charges}

We actually focus on the Euclidean counterparts of $sl(N,\mathbb{R})
\oplus sl(N,\mathbb{R})$ higher-spin theories, which are described by
Chern-Simons actions with gauge algebra $sl(N,\mathbb{C})$
(analogously to the description of three-dimensional Euclidean gravity
by a $sl(2,\mathbb{C})$ Chern-Simons theory as reviewed e.g.\
in~\cite{review_banados}). This is indeed the setup that simplifies
the comparison with the minimal model spectrum \cite{conical}. 

The bulk field equations are solved by any flat
$sl(N,\mathbb{C})$-valued connection. However, as the Chern-Simons theory is
defined on a manifold with boundary, one also has to impose suitable
boundary conditions on these connections. To describe them it is
convenient to introduce a radial coordinate $r$ and to parameterise a
cylinder at surfaces of constant $r$ by the complex coordinates
$w,\bar{w}$. They are defined by $w = \phi + i t_E$, where $t_E$ is
the Euclidean time and $\phi$ is an angular coordinate with
periodicity $\phi \sim \phi + 2\p$. In \cite{HR,spin3,GH,Wlambda} it was proposed to select asymptotically AdS solutions by requiring that one can cast their connections in the form
\begin{align}
& A(r,w,\bar{w}) \,=\, b^{-1} a(w) b\, dw + b^{-1}db \ , \\
& a(w) \,=\, J_{1} + \sum_{j\,=\,2}^{N} \big(-\sqrt{k}\big)^{-j}\, u_{j}(w)\, e_{1,\,j} \ , \label{u-gauge}
\end{align}
where $b = b(r)$ is a generic group-valued element depending only on the
radial coordinate, $J_1$ is the matrix $(J_1)_a{}^b = -\,
\d_a{}^{b+1}$ and $e_{i,\,j}$ is the matrix with one entry $1$ in the
$i^{\text{th}}$ row and $j^{\text{th}}$ column, and zeroes
otherwise.\footnote{In \cite{HR,spin3} the connections were actually
presented in a different form that is related to \eqref{u-gauge} by a gauge transformation (see e.g.\ sect.~4 of \cite{Wlambda}).}
We denote by $k$ the level of the Chern-Simons theory.\footnote{We follow
the conventions of~\cite{conical} where the trace is normalised such
that it satisfies~$\text{tr}\,J_{a}J_{b}=\binom{N+1}{3}\kappa_{ab}$
for the $sl(2)$ generators $J_{a}$ and the standard $sl(2)$ Killing
form $\kappa_{ab}$. In this convention the central charge of the
asymptotic Virasoro subalgebra is $c=N (N^{2}-1)k$.} 

The conical defect solutions of \cite{conical} further preserve
time-translation and rotational invariance, and this is achieved by
considering only constant $u_j$. Moreover, the solutions are smooth in
the sense that the holonomy around the contractible $\phi$-circle is
trivial. This signals the absence of singularities in the gauge
configuration. When this condition is satisfied, the matrix
\eqref{u-gauge} can be diagonalised with imaginary eigenvalues
\begin{equation}
i\,n_{j}' \,=\, i \Big(m_{j}-\frac{\sum_{j}m_{j}}{N}\Big) \ ,
\end{equation}
where $m_j$ are integers satisfying $m_1 < \cdots < m_N$
\cite{conical}. Comparing the characteristic polynomial for $a$ in the
$u$-basis~\eqref{u-gauge} and in the diagonal basis one
concludes
\begin{equation}\label{conicalcharge}
u_{j} \,=\,  \big(-i\sqrt{k}\big)^{j} P_{j} (n_{1}',\dotsc ,n_{N}') \ ,
\end{equation}
where $P_{j} (x_{1},\dotsc ,x_{N})$ denotes the $j^{\text{th}}$ elementary symmetric polynomial in the variables $x_{1},\dotsc ,x_{N}$:
\begin{equation} \label{symm_pol}
P_j(x_{1},\dotsc ,x_{N}) \,=\, \sum_{1\leq k_1<\cdots<k_j\leq N} x_{k_1} x_{k_2} \cdots x_{k_j} \, .
\end{equation}
Each conical solution is thus specified by a list of $N$ distinct integers, and its $u$-charges are easily expressed in terms of them through \eqref{conicalcharge}.

\section{Charges from Miura transformation}

When one restricts the connections of the Chern-Simons theory to those
that correspond to asymptotically AdS solutions of the
form~\eqref{u-gauge}, one finds an asymptotic W$_{N}$ symmetry algebra
on the boundary that is generated by the boundary currents $u_{j}$ \cite{HR,spin3,GH,Wlambda}:
this is the classical Drinfeld-Sokolov reduction in the 'u-gauge'.
The quantum counterparts $U_{j}$ of the fields $u_{j}$ can be obtained
from the quantum Drinfeld-Sokolov reduction of $sl(N)$ (see e.g.\
\cite{W-review}) which defines these higher-spin currents in terms of
normal-ordered products of free fields by the so-called quantum Miura
transformation~\cite{Lukyanov},
\begin{equation}\label{quantumMiura}
(i\alpha_{0}\partial)^{N} + \sum_{j=2}^{N}U_{j} (z)
(i\alpha_{0}\partial)^{N-j} \,=\,
\big((i\alpha_{0}\partial -i\epsilon_{1}\cdot J (z))\dotsb 
(i\alpha_{0}\partial -i \epsilon_{N}\cdot J (z)) \big) \ ,
\end{equation}
where $J (z)=i\partial \vf (z)$ is a spin-1 current taking
values in the weight space of $sl(N)$ (see e.g.\ sect.~6.3.3 of
\cite{W-review}).
The $\epsilon_{i}$ are the weights of the fundamental representation;
$\epsilon_{1}=\omega_{1}$ is the first fundamental weight, and
$\epsilon_{i+1}=\epsilon_{i}-\alpha_{i}$, where $\alpha_{i}$ are the
simple roots. We use the standard normalisation where 
\begin{equation}
\epsilon_{i}\cdot \epsilon_{j} \,=\, \delta_{ij} - \frac{1}{N} \ .
\end{equation}
The parameter $\alpha_{0}$ is related to the central charge of the
$W_{N}$ algebra by
\begin{equation}\label{centralcharge}
c (N,\alpha_{0}) \,=\, (N-1)\big(1-N (N+1)\,\alpha_{0}^{2} \big) \ .
\end{equation}
The $U_{j}$ are holomorphic fields of scaling weight $j$.
They are not primary fields, so they transform non-trivially under
conformal transformations. On the other hand, this transformation can be derived from the transformation property of the free spin-1
current~$J$. 

In the above free-field construction there is a background charge for
the free field. In particular the energy-momentum tensor has a shift
proportional to the derivative of the current $J$,
\begin{equation}\label{energymomentum}
U_{2} (z) \,=\, T (z) \,=\, \frac{1}{2} (J\cdot J) (z) -\alpha_{0} \rho \cdot
\partial J (z) \ ,
\end{equation}
where $\rho \,=\, - \sum_{j} j\, \e_j$ is the Weyl vector. Under a conformal transformation
$z\mapsto w (z)$ the free field transforms as
\begin{equation}
J (z) \to \tilde{J} (w) \,=\, \frac{1}{w' (z)}\bigg(J(z) -\alpha_{0}\rho
\frac{w''(z)}{w'(z)}  \bigg) \ .
\end{equation}
From this rule one can now deduce the transformation
property of the fields $U_{k}$. There is however one subtlety: the
$U_{k}$ are determined in terms of normal-ordered products. Under a
general conformal transformation the notion of normal-ordering
changes, and this can lead to additional terms in the transformation
as we will briefly discuss at the end of this section.

We are primarily interested in the transformation from the plane to
the cylinder where we can compare the results to the higher-spin
computation in the bulk. If $z$ is the coordinate on the plane, the
relation to the cylinder with coordinate $w$ is $z=e^{-iw}$. The spin-1 currents then transform as
\begin{equation}\label{freefield}
J (z) \to \tilde{J} (w) \,=\, -\,i \big(zJ (z) + \alpha_{0} \rho \big) \ .
\end{equation}
If we ignore the normal-ordering subtleties (which do not play a
role in the semi-classical limit in which we compare the charges), the
transformed fields $\tilde{U}_{j}$ on the cylinder are determined by
\begin{multline}
(i\alpha_{0}\partial_{w})^{N} + \sum_{j=2}^{N}\tilde{U}_{j} (w) (i\alpha_{0}\partial_{w})^{N-j} \\
=\,  \Big(\big(i\alpha_{0}\partial_{w} - \epsilon_{1}\cdot ( z J
(z)+\alpha_{0}\rho )\big)\dotsb 
\big(i\alpha_{0}\partial_{w} - \epsilon_{N}\cdot( z J
(z)+\alpha_{0}\rho ) \big) \Big) \ .
\end{multline}
In particular we can now determine the eigenvalues of the zero
modes of $\tilde{U}_{j}$ on a highest-weight representation: let
$|\Lambda \rangle$ be a highest-weight state for the free field with 
\begin{equation}
J_{0}|\Lambda \rangle \,=\, \Lambda |\Lambda \rangle \ .
\end{equation}
We then have
\begin{equation}
(i\alpha_{0}\partial_{w})^{N} + \sum_{j=2}^{N}\tilde{U}_{j,0} (i\alpha_{0}\partial_{w})^{N-j}
\,=\, \Big(\big(i\alpha_{0}\partial_{w} - \epsilon_{1}\cdot ( \Lambda +\alpha_{0}\rho )\big)\dotsb 
\big(i\alpha_{0}\partial_{w} - \epsilon_{N}\cdot( \Lambda
+\alpha_{0}\rho ) \big) \Big) \ , 
\end{equation}
and hence
\begin{equation}\label{charge}
\tilde{U}_{j,0} \,=\,  (-1)^{j}\, P_{j} \big(\epsilon_{1}\cdot ( \Lambda
+\alpha_{0}\rho ),\dotsc ,\epsilon_{N}\cdot( \Lambda +\alpha_{0}\rho)
\big)\ ,
\end{equation}
where $P_{j}$ is the $j^{\text{th}}$ elementary symmetric polynomial, see~\eqref{symm_pol}.

Before we compare these values to the charges of the conical solutions
in the next section, we want to briefly discuss the quantum
corrections to the charge formula~\eqref{charge}. Let us illustrate
this with the well-known example of the energy-momentum tensor 
$U_{2}(z)=T (z)$ (see~\eqref{energymomentum}). Going to the cylinder
we obtain the transformed field
\begin{equation}
\tilde{U}_{2} (w) \, =\, \frac{1}{2}\big(\tilde{J}\cdot\tilde{J} \big)_{w} (w)
-\alpha_{0}\rho \cdot \partial_{w}\tilde{J} (w)\ ,
\end{equation}
where the subscript $_{w}$ on the parentheses signals that the normal
ordering is taken on the cylinder. Replacing $\tilde{J}$ according
to~\eqref{freefield} we find
\begin{equation}
\tilde{U}_{2} (w) \, =\, -z^{2} \bigg( \frac{1}{2}\big(J\cdot J\big)_{w} (z)
-\alpha_{0}\rho \cdot \partial_{z}J (z)\bigg)
-\frac{1}{2}\alpha_{0}^{2}\rho^{2}\ .
\end{equation}
The normal ordering defined via the subtraction of the singular
terms in the operator product expansion gives a different result on
the cylinder and on the plane, with the relation
\begin{equation}
\big(J\cdot J \big)_{w} (z) \, = \, \big(J\cdot J \big)_{z} (z)
-\frac{N-1}{12\,z^{2}} \ .
\end{equation} 
Therefore we find the familiar transformation
\begin{equation}
\tilde{U}_{2} (w) \, =\, -\,z^{2} U_{2} (z) +\frac{N-1}{24}
-\frac{1}{2}\alpha_{0}^{2}\rho^{2}\, =\, -\,z^{2} U_{2} (z) +\frac{c}{24}\ .
\end{equation}
In the semi-classical limit where $c\to\infty$ (and $N$ is fixed) we
observe that the normal-ordering shift $\frac{N-1}{24}$ (which does not grow with $c$)
is a subleading correction as expected. 
These shifts become less trivial for higher-spin
charges: for example the transformation of the spin-4 field is given by
\begin{align}
\tilde{U}_{4} =\, & z^{4}\,U_{4} + 3\,i\,\frac{N-3}{2}\,\alpha_{0}\,z^{3}\,U_{3}
+\frac{(N-2) (N-3)}{24}\bigg(\alpha_{0}^{2}\, (N-13) +\bigg\{
-\frac{1}{N}\bigg\}\bigg) z^{2}\,U_{2} \nonumber\\
& -\frac{1}{192}\, \binom{N-1}{3} \bigg(\alpha_{0}^{4}\,N \,(N+1)
(N+\tfrac{7}{5}) +\bigg\{ -\alpha_{0}^{2}\,\frac{142}{5} + 2\,\alpha_{0}\,
(N-13) + \frac{1}{N}\bigg\}\bigg) \ ,
\end{align}
where the quantum corrections due to normal ordering are contained in curly brackets. It would be interesting to work out the quantum
corrections in general, since they would allow one to test possible proposals for a quantum description of higher-spin theories in the bulk.
 
\section{Comparison and discussion}

Degenerate representations of the $W_{N}$ algebra are labelled by two
$sl(N)$ highest weights $(\Lambda_{+},\Lambda_{-})$ \cite{W-review}, and the
corresponding weight vector $\Lambda$ is given by 
\begin{equation}
\Lambda \,=\, \alpha_{+}\Lambda_{+} + \alpha_{-}\Lambda_{-} \ ,
\end{equation}
where $\alpha_{+}\alpha_{-}=-1$ and
$\alpha_{+}+\alpha_{-}=\alpha_{0}$. According to the proposal
in~\cite{conical-CFT} conical solutions correspond in the
semi-classical limit ($c\to\infty$ with $N$ fixed) to representations
$(0,\Lambda_{-})$, whereas the representations $(\Lambda_{+},\Lambda_{-})$
describe excitations of the scalar in the Vasiliev theory on the conical background labelled by $(0,\Lambda_{-})$.
We can take the semi-classical limit by giving $\alpha_{-}$ a large imaginary value, $\alpha_{-}\approx i \frac{c}{N (N^{2}-1)}$. In this limit $\a_+$ vanishes since it scales as $c^{-1}$: as a result the charges \eqref{charge} do not depend on $\L_+$ and become
\begin{equation}
\tilde{U}_{j,0} \,=\, ( -\alpha_{-})^{j} P_{j} \big(\epsilon_{1}\cdot ( \Lambda_{-}
+\rho ),\dotsc ,\epsilon_{N}\cdot( \Lambda_{-} +\rho)
\big) \ + \ \dotsb \ .
\end{equation}
Upon identifying $\alpha_{-} \approx i\sqrt{k}$ (which gives the
correct matching for the central charge) and $\epsilon_{i}\cdot
(\Lambda_{-}+\rho) = n_{i}'$ we find precise agreement
with~\eqref{conicalcharge}. 

Therefore we have shown that the spectrum of conical solutions exactly
matches the spectrum in the conformal field theory in the
semi-classical limit, providing an important check of the
proposed higher-spin AdS$_{3}$/CFT$_{2}$ duality.\footnote{Further evidence for
the correct identification of the conical solutions comes from a
matching of the four-point function in the CFT with a two-point
function in the background of a conical solution in the
bulk~\cite{four-point}.}

The construction of smooth asymptotically AdS connections was generalised to
higher-spin theories based on the infinite-dimensional gauge algebra
hs$(\lambda)$ in~\cite{conical-lambda}. A class of these solutions can
be seen as a continuation of solutions in the $sl(N)$ theories. The
charge formulas that we obtained in this note can be straightforwardly
continued to this case: for a fixed representation $\Lambda_{-}$ the
charges stabilise for large $N$ to rational functions in $N$, and by
replacing $N\to\lambda$ we obtain the charge for the hs$(\lambda)$
theory.  

\section*{Acknowledgements}

We thank M.R.~Gaberdiel, T.~Proch\'azka and J.~Raeymaekers for useful discussions. We also thank the Galileo Galilei
Institute for Theoretical Physics for the hospitality and the INFN for
partial support during the program on 'Higher Spins, Strings and
Duality' where this project was started. The work of AC is partially supported by IISN - Belgium (conventions 4.4511.06 and 4.4514.08), by the ``Communaut\'e Fran\c{c}aise de Belgique" through the ARC program and by the ERC through the ``SyDuGraM" Advanced Grant.


\end{document}